\documentclass{sig-alternate}

\usepackage{booktabs}
\usepackage{dsfont}
\usepackage[hidelinks]{hyperref}
\usepackage{framed}
\usepackage{tikz}
\usepackage[font={it},skip=5pt]{caption}

\makeatletter
\def\@copyrightspace{\relax}
\makeatother

\begin{document}

\title{Event Search and Analytics}
\subtitle{Detecting Events in Semantically Annotated Corpora for Search \& Analytics}

\numberofauthors{1}
\author{
\alignauthor
       Dhruv Gupta\\
       \affaddr{Max Planck Institute for Informatics, Saarbr\"{u}cken, Germany}\\
       \affaddr{Saarbr\"{u}cken Graduate School of Computer Science, Saarbr\"{u}cken, Germany}\\
       \email{dhgupta@mpi-inf.mpg.de}
}

\maketitle
\begin{abstract}
 In this article, I present the questions that I seek to answer in my PhD research. 
 I posit to analyze natural language text with the help of semantic annotations and 
 mine important events for navigating large text corpora. Semantic annotations such 
 as named entities, geographic locations, and temporal expressions can help us mine 
 events from the given corpora. These events thus provide us with useful 
 means to discover the locked knowledge  in them. I pose three problems that can help 
 unlock this knowledge vault in semantically  annotated text corpora: \emph{i.} identifying important
 events; \emph{ii.} semantic search; and \emph{iii.} event analytics.
 
\end{abstract}

\keywords{Information Retrieval; Text Analytics; Text Summarization;
Semantic Annotations; Diversity \& Novelty; Semantic Search}

\section{Introduction}

Information retrieval systems have largely relied 
on word statistics in text corpora to satisfy 
information needs of users by retrieving documents 
with high relevance for a given keyword query. In my PhD 
research I hypothesize that information needs of users can 
be satisfied to a greater extent by using 
\emph{events} as a means of navigating text corpora. 
Events in our context would be an act performed by certain 
actor(s) at a specific location during a specific time 
interval. An example would be : \emph{Usain Bolt 
won the gold medal at the 2008 summer Olympics in 
Bejing}. With the availability of annotators that 
can provide us with accurate semantic annotations in form of 
named entities, geographic locations, and
temporal expressions; we can leverage the growing number of knowledge 
resources such as Wikipedia~\cite{wiki}
and ontologies such as Freebase~\cite{freebase} 
to understand natural language text and mine important events. 
Formally the central hypothesis can be stated as follows:
\begin{figure}[h]
 \begin{framed}
  \vspace*{25pt}
  {\textbf{Central Hypothesis}
   Given text corpora with semantic annotations; traditional 
   information retrieval models can be improved by utilizing 
   knowledge about \emph{events} and using \emph{events} as 
   proxies for information need.}
  \vspace*{25pt}
  \end{framed} 

\end{figure}

As a toy example consider the following text 
snippet~\footnote{\scriptsize
\url{http://www.bbc.com/sport/0/athletics/34032366}} with demonstrative 
semantic annotations in Figure~\ref{fig:text} :

\begin{figure}[h]
  \begin{framed}
  \vspace*{25pt}
  {$\ldots$ \underline{Beijing}{\scriptsize${\langle{Wiki:Beijing}\rangle 
  \langle{Geo:(39.55, 116.23)}\rangle}$} 
  where \underline{Bolt} {\scriptsize${\langle{Wiki:Usain\_Bolt}\rangle}$} 
  announced himself to the world with two Olympic golds and two world records in 
  \underline{2008}{\scriptsize${\langle{Time:[01-01-2008, 31-12-2008]}\rangle}$} $\ldots$}
  \vspace*{25pt}
  \end{framed} 
  \caption{Semantically annotated text~$^1$ snippet}
  \label{fig:text}
\end{figure}

In the text snippet (Figure~\ref{fig:text}), we 
obtain the named entity \textsf{Usain Bolt} whose 
mention has been identified and disambiguated 
to point to an external knowledge source.
Also identified is a geographical location - 
\textsf{Beijing}, which is disambiguated
and resolved to its geographical coordinates. 
Likewise the temporal expression \textsf{2008} has also been 
resolved to time range.
Having these semantic annotations 
we can now devise algorithms that can deduce that 
the event is that of Usain Bolt winning
Olympic competition in Beijing, China.

The goal of the proposed research is to 
leverage the semantic annotations for mining
important events and use them to navigate text 
corpora. The research will find application 
in many domains of research such as \emph{digital 
humanities}, in which social scientists 
are interested in computational history in large 
digital-born text collections. 
Anthropologists are interested in cultural 
and linguistic shifts that occur in such 
collections. Collectively we can allow 
\emph{computational 
culturomics}~\cite{culturomics} 
on corpora to study cultural
trends. Events can also be used to link 
information in multiple and diverse text 
collections. 
In short, important events provide a way to 
create a gist from semantically annotated corpora, 
which otherwise is not possible through manual
human effort.

\textbf{Outline}. The article consists of: 
\begin{itemize}
	\item a literature survey (Section~\ref{sec:background});
	\item an overview of the research problems (Section~\ref{sec:problem}); 
	\item available corpora, test sources and  evaluation measures for research 
		 (Section~\ref{sec:evaluation}); 
	\item discussion of few open technical problems (Section~\ref{sec:discussion}).
\end{itemize}


\section{Related Work}~\label{sec:background}

 In this section I discuss the progress already made in the area of analyzing different semantic annotations in isolation as well as 
 in conjunction for some of the problems proposed.

\textbf{Temporal Information Retrieval and Extraction}. Researchers have considered only temporal annotations in text corpora to improve
retrieval effectiveness by analyzing the time sensitivity of keyword queries and incorporating the time dimension in retrieval models.
Some methods of analysis of time-sensitive queries rely on publication dates of documents~\cite{diaz_profile,nattiya_2010}, while others
also look at the temporal expressions in document contents~\cite{dhruv_2014}. Several works also take into account the time dimension
for re-ranking documents~\cite{klaus_2010} and diversifying them along time~\cite{klaus_2013, nattiya_2014}. One of the seminal works in
extracting temporal events was by Ling and Weld~\cite{ling_tie}. They outline a probabilistic model to solve the 
problem of extracting relations from text with temporal constraints.

\textbf{Important Events in Annotated Corpora}. One of the most important seminal works in identifying existing and emerging events were the 
various tasks in \emph{Topic Detection and Tracking} (TDT)~\cite{tdt_book}.
The TDT program aimed to ``search, organize and structure'' broadcast news media from multiple 
sources. The five tasks laid within the ambit of TDT where topic tracking,
link detection, topic detection, first story detection,  and story segmentation.
The task of topic tracking required to build a system to detect \emph{on topic stories} 
from an evaluation corpus after being trained on a set \emph{on topic} stories.
The link detection task involved answering 
a boolean query to whether two given \emph{stories} are related by a common topic.
The topic detection task comprised of declaring new \emph{topics} from 
incoming \emph{stories} which had not been presented to the system. 
First story detection was another boolean decision task 
of determining whether the given \emph{story} is a seed story (first-story) to create a 
new \emph{topic} cluster. Story segmentation task required segmentation of
an incoming stream of text into \emph{stories}.

Focusing specifically on extracting and summarizing events in the future, Jatowt and Yeung~\cite{adam_cikm11} 
present a model-based clustering algorithm. The clustering considers both textual and temporal similarities.
For computing temporal similarity, the authors model time as a probability distribution by utilizing different
family of distributions based on whether its is singular time point, starting date or and ending date. The 
similarity is then computed using KL-divergence. 

Radinsky et al.~\cite{kira_jair} present an algorithm \emph{Pundit}, which based on the past events in text 
is able to predict a future event given a query to the system. The events are represented as multidimensional 
attributes such as time, geographic location and participating entities. The algorithm derives these events 
from external text collection and builds an \emph{abstraction tree}, which is the result from
hierarchical agglomerative clustering. In order to predict the future \emph{Pundit} is trained to select
the most similar cluster from the \emph{abstraction tree} and produce an event representation.

The work by Yeung and Jatowt~\cite{yeung_cikm11} tackles the problem of analysis of historical events in multiple 
large document collections. They utilize \emph{latent Dirichlet allocation} to identify \emph{topic} distributions
along time. Thereafter they perform analytics to answer questions such as \emph{i.} significant years and topics, \emph{ii.}
triggers that caused remembrance of the past and \emph{iii.} historical similarity of countries.

Most recently, Abujabal and Berberich~\cite{mppf} present a system which identifies important events in 
text collections by counting frequent itemsets of sentences containing named entities and temporal 
expressions. For evaluation they resort to \emph{Wikipedia's} event directory as a ground truth.

\textbf{Semantic Search}. Summarizing text collections in a timeline visualization is a natural choice.
Swan and Allan~\cite{swan_timeline} present an approach for producing a timeline that depicts most important topics and events closely modeled on the 
\emph{Topic Detection and Tracking} task. The algorithms analyzes features based on named entities and noun phrases.
The analysis involves construction of $2 \times 2$ contingency table on presence or absence of features, and subsequent 
measurement of $\chi^2$ statistic for measuring significance of co-occurrence of a pair of features. 

The seminal work by Baeza-Yates~\cite{yates_future} proposed a \emph{future retrieval} (FR) system. The FR system 
considers both text and temporal expressions to identify future events that might be relevant to an input query. Baeza-Yates 
outlined the components of a FR system to be composed of an \emph{information extraction} (IE) module, \emph{information
retrieval} (IR) module, and  a \emph{text mining} (TM) module. The IE module would act as a temporal annotator; identify temporal expressions
and normalize them. The IR system is designed to incorporate the time dimension in an index; thus retrieving documents
with text and time similarity. The TM module would identify the most relevant topics given a time period. 
He presented a retrieval model, in which each document
consists of a multiple temporal events. A temporal event consists of a time segment and its associated likelihood of occurring. 
The score of the document is thus obtained by its 
textual similarity and the maximum likelihood of all the temporal events in that document. 

Bast and Buchhold~\cite{bast_index} outline a joint index structure over ontologies and text.
Which allows for fast semantic search and provide context sensitive auto-complete suggestions.  

Events as a means of search document collections has also been explored by 
Str\"{o}tgen and Gertz~\cite{jannik_event}. Events were modeled by the geographic location and time of their occurrence. 
For temporal queries expressed in simple natural language they outline an extended Backus-Naur form (EBNF) language that incorporates time intervals 
with standard boolean operations. Geographical queries are also modeled as EBNF language, however the input for them 
is a minimum bounding rectangle (MBR). Using this multidimensional querying model the user is able to visualize
search results in form of events; which are additionally represented on a map.

Giving special attention to geographical information retrieval, Samet et al.~\cite{samet_news} present a system
\emph{NewsStand}, that is able to resolve and pinpoint a news article based on the geographic information present 
in its content. They discuss various methods for toponymn resolution, which is in essence disambiguating the geographic
location based on its surface form in the news content. The system involves a streaming clustering algorithm that can
keep track of emerging news in new locations and present them in a map-based interface.

\begin{figure*}[t]

	\small

	\centering

	\begin{tabular}{l|p{4cm}p{4cm}p{4cm}}		

	\toprule

	\multicolumn{4}{l}{\textbf{Query:} \texttt{ summer olympics}}\\

	\midrule

	\textbf{Event} & $c_1$ & $c_2$ & $c_3$ \\

	\midrule

	\textbf{Words}  & micheal, phelps, bejing, china, tibet & london, usain, bolt, england, badminton & rio, brazil,copacabana, deodoro, maracan\~a \\

	\textbf{Time} &$[08-08-2008, 24-08-2008]$ & $[27-07-2012, 12-08-2012]$ & $[05-08-2016, 21-08-2016]$ \\

	\textbf{Location} & $\langle Beijing, China \rangle$ & $\langle London, England \rangle$ & $\langle Rio de Janeiro, Brazil \rangle$ \\

	\textbf{Entities} & $\langle China \rangle$, $\langle Micheal\_Phelps \rangle$ & $\langle England \rangle$, $\langle Badminton \rangle$ &  $\langle Brazil \rangle$, $\langle Copacabana \rangle$ \\

	\bottomrule

	\end{tabular}

	\caption{Example important events for an example query \texttt{summer olympics}}

	\label{fig:event}

\end{figure*}

\textbf{Event Analytics}. By disambiguating and linking named entities to ontologies, Hoffart et al.~\cite{aesthetics,stics} provide a framework 
for semantic search and performing analytics on them. They provide features for giving auto-complete suggestions in the form of similar 
entities for the input named-entity. In~\cite{aesthetics} they provide analytics that leverage accurate entity counts and 
provide entity co-occurrence statistics which is helpful in analyzing semantically similar named-entities.

\section{Research Objectives}~\label{sec:problem}
Given the text corpora with semantic annotations, I describe three 
important research problems in this section: \emph{i.} identifying important events; 
\emph{ii.} using identified events for improving retrieval effectiveness; and 
\emph{iii.} using identified events for analytics.

\subsection{Notation}
Let us consider multiple corpora for the purpose of 
analysis. This allows us to capture frequently 
occurring events as well as link similar events 
across corpus. Given corpora
$$
D = \bigcup_{k=1}^{N} D_k,
$$
where each document $d \in D$ consists of word sequence $x$
at appropriate granularity (e.g. paragraph or sentence):
$$
  d = \bigcup_{i=1}^{n} x_i.
$$
Further each $x \in d$ contains semantic annotations in
form of \emph{i.} named entities ($e$), \emph{ii.} geographical location ($g$),
and \emph{iii.} temporal expressions ($t$). Additionally $x$ also
consists of the a bag of words $\mathcal{W}$ drawn from 
a vocabulary $\mathcal{V}$. Formally represented as:
$$
  x = \langle \mathcal{E}, g, t, \mathcal{W} \rangle
$$

\subsection{Problem Definition}

The objective is to design a family of algorithms:
$$
\textsc{Event*}(X,Q,\alpha)
$$ 

where $X = \bigcup x$, $Q$ represents an input query
and $\alpha \in \mathds{R}^m$, where $\alpha$ is set of parameters.

The input query $Q$ can be a combination of following input components:
\emph{i.} keyword query $q$, \emph{ii.} time $q_{time}$, \emph{iii.} geographical location $q_{geo}$,
and \emph{iv.} named entity $q_{entity}$.

Given the input, we need to design the algorithms $\textsc{Event*}$ according to the
different problems. We discuss the design objectives for the three different purposes in this section.

\textbf{Identifying Important Events}. \emph{Events} are the proposed building blocks for 
further text analysis. An \emph{event} in our context is defined to be an activity or an act
involving  named entities that happens in a specific geographical location anchored to a specific time
interval. Mathematically, given a multidimensional query :
$$
Q = \langle q, q_{time}, q_{geo}, g_{entity} \rangle,
$$ 
and a subset of highly 
relevant documents $R \subseteq D$, the algorithm for this purpose $\textsc{EventDetect}$
should produce a set of ordered events :
$$
 \mathcal{C} = \{ c_1, c_2, \ldots, c_k \},
$$

where, $c = \langle \mathcal{E}, g, t, \mathcal{W} \rangle$. The event $c$ is 
hence described by the participating named entities $\mathcal{E}$,
its location $g$, its time of occurrence $t$, and frequently occurring 
contextual terms around these semantic annotations $\mathcal{W}$.
This requires proposing a probability mass function, $P(\mathcal{C}, R)$, 
using which we can impose a total order on $\mathcal{C}$. 

As an example consider the keyword-only query \texttt{summer olympics}
to the processed corpora of news articles. The designed algorithm
shall then identify the important events as in Figure~\ref{fig:event}.

\textbf{Diversifying and Summarizing Search Results} are retrieval tasks
that try to address the information need underlying an ambiguous query at 
different levels of textual granularity. Each task tries to maximize 
the coverage of different information needs underlying the given ambiguous 
query. As information intents, we propose to use the mined
set of \emph{events}. Accomplishing these tasks would 
allow for automatic creation of \emph{event timelines} or \emph{entity biographies}. 
We briefly discuss an intuition of achieving the same.

When diversifying search results we would like to present users with 
\emph{documents} such that the user finds \emph{at least} one document 
that satisfies her information intent. For this we need to devise 
an algorithm $\textsc{EventDiverse}$ which considers
as an input $Q$ and $R \subseteq D$. As an output it returns
a set of documents $S \subset R$ which cover all events in $\mathcal{C}$.

Summarizing search results would require us to construct an 
algorithm $\textsc{EventSummary}$ to piece together,
\emph{sentences} $\hat{S} = \bigcup x$, such that the
text summary covers all events in $\mathcal{C}$.

\begin{figure}[t]
 \centering
 \includegraphics[scale=0.75]{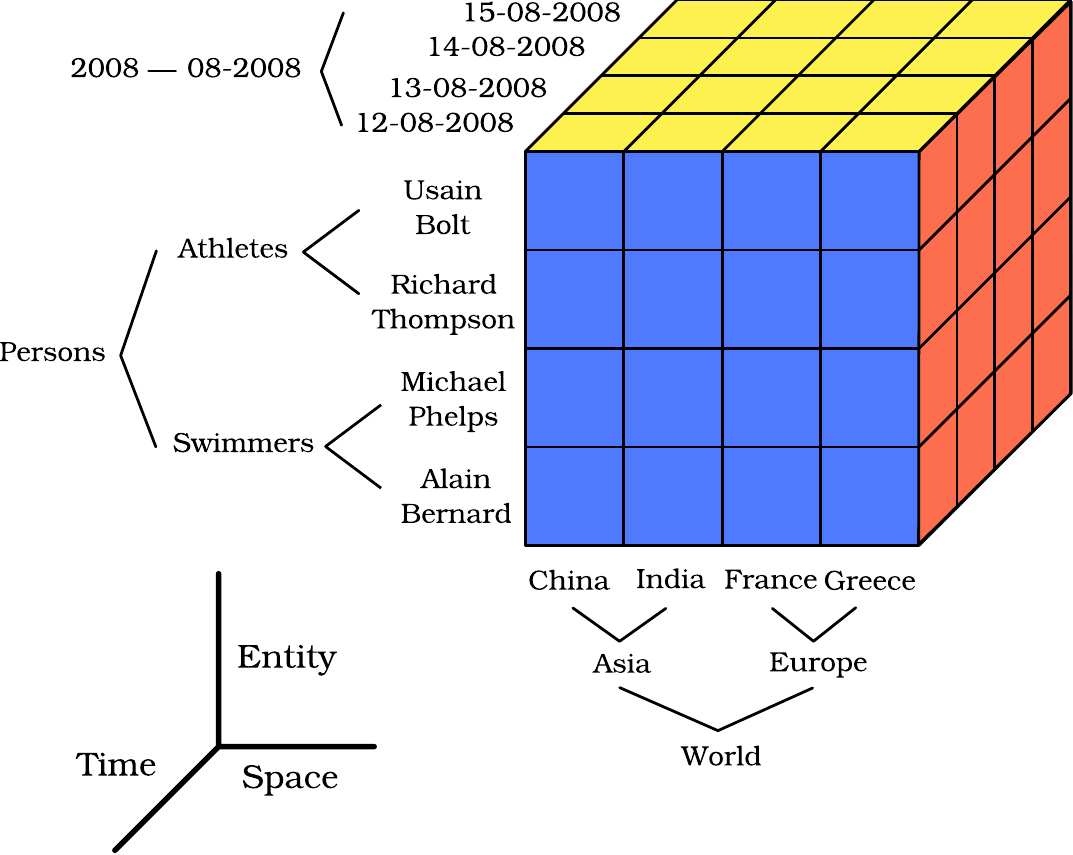}
 \caption{Example data cube based on set of events $\mathcal{C}$}
 \label{fig:data_cube}
\end{figure}

\textbf{Semantic Search and Analytics}. The mined set of \emph{events} can 
further be utilized for search and analytics. For this purpose we can utilize
inherent hierarchy in the semantic annotations. For example a given year \textsf{2015}
can be broken down to different months and subsequently days in those months.
Similarly, we can utilize the \emph{type hierarchies} in named entities. Such
as \textsf{Usain Bolt} and \textsf{Justin Gatlin} are subtypes of \textsf{Athletes}.
This can jointly be modeled by using the concept of a \emph{data cube}~\cite{han_dm}
as shown in Figure~\ref{fig:data_cube}.

Formally, given a query $Q$, the objective would to first model 
the mined set of events as a \emph{data cube} and subsequently provide 
\emph{data cube operations}~\cite{han_dm}: 

\begin{itemize} \setlength{\itemsep}{-1pt}
 \item roll ($\bigcirc$), 
 \item slice ($\ominus$),
 \item dice ($\oplus$),
 \item drill up ($\bigtriangleup$),
 \item drill down ($\bigtriangledown$).
\end{itemize}

\begin{figure}[t]
 \centering
 \includegraphics[scale=0.50]{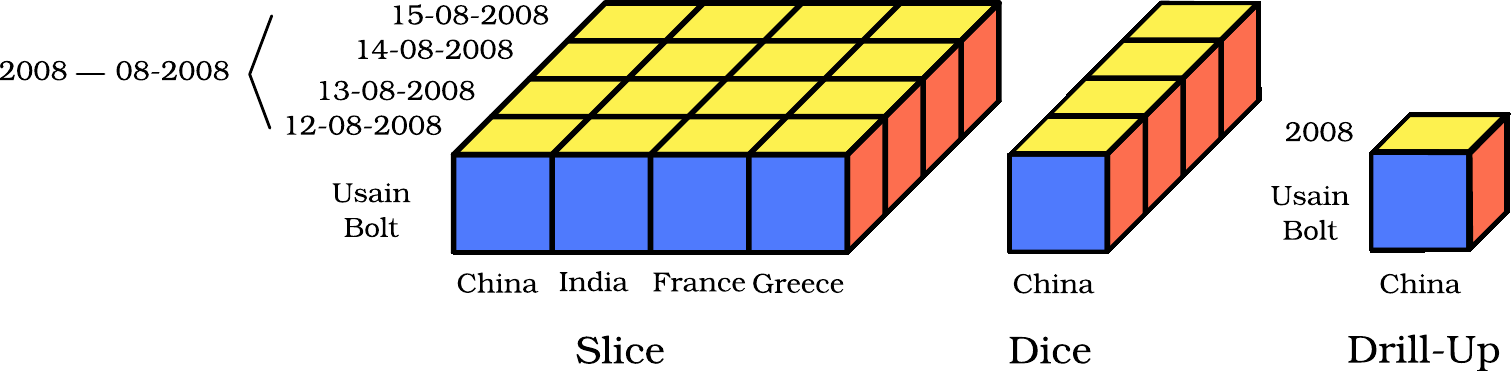}
 \caption{Example data cube operations for the query \texttt{all races won during 2008 by usain bolt in china} }
 \label{fig:cube_opr}
 \vspace{-10pt}
\end{figure}

As a concrete example consider 
the query \texttt{all races won during 2008 by usain bolt in china}. 
To produce an appropriate result the sequence of operations would be:
first a slice on the entity \textsf{Usain Bolt}; second dice on 
\textsf{China}; and finally drill up to year \textsf{2008} (see Figure~\ref{fig:cube_opr}).
\section{Data}

\textbf{Corpora}. There are several readily available massive data sets. They are 
 available from news corporation such as the \emph{New York Times}~\cite{nyt}, 
 \emph{English Gigaword}~\cite{gigaword}. These corpora have the benefit of 
 being available with reliable publication dates and grammatically well-formed
 text. On larger scale are Web collections such as \emph{ClueWeb'09}~\cite{clueweb09}/'12~\cite{clueweb12},
 which are not always accompanied by reliable creation dates and many are ill-formed documents.

\textbf{Semantic Annotations}. The text corpora next need to be annotated for text mining.
I explain how to obtain the different semantic annotations in the following paragraphs.

\textbf{Named Entities}. For disambiguating and linking named entities in text to an external knowledge 
source such as \emph{Wikipedia} \cite{wiki} or an ontology such as YAGO~\cite{yago} or Freebase~\cite{freebase};
I use the AIDA system~\cite{aida}. The AIDA system does named entity disambiguation and linking
by leveraging contexts extracted from ontologies such as YAGO. For Web collections such ClueWeb'09/'12 
the entity disambiguation and linking has been released as \emph{facc1 : Freebase annotation of ClueWeb
Corpora}~\cite{facc}.

\textbf{Geographical Locations} can be obtained by utilizing \emph{geographic} named entities
such as those known to be cities, countries, or continents.
Geographical relations stored in an ontology can be used to resolve these locations to its 
geographical coordinates. Having obtained a set of coordinates, we can subsequently construct 
a geographical representation such as a \emph{minimum bounding rectangle} over the coordinates.

\textbf{Temporal Expressions}, both implicit and explicit, can be extracted and  
normalized from text by using \emph{temporal taggers} such as HeidelTime~\cite{heideltime}
or SUTime~\cite{sutime}.

\section{Evaluation}~\label{sec:evaluation}

To test our approach we need to construct query sets that contain an event description associated with the query; 
along with participating named entities, geographical locations where the event took place and relevant time interval
associated with it. I describe a tentative approach to achieve this here.
	
\textbf{Test Data}. To evaluate the correctness of the various algorithms, I plan to use reliable encyclopedic 
resources on the Web such as \emph{Wikipedia}~\cite{wiki} or other curated knowledge sources. For 
an objective evaluation, I propose the following different sources depending on the algorithm under evaluation.
\begin{itemize} 
  \item Identify important events 
  \begin{itemize}  
    \item Events in a particular year/decade etc. pages available on \emph{Wikipedia}~\cite{wiki_year}.
    \item Testing of past events can be done by extracting important topics from \emph{Category} pages on various 
	  historical topics on \emph{Wikipedia}~\cite{wiki_past}.
    \item Events in the future can be evaluated by using important infrastructure projects, engineering
          projects etc. These can be extracted from \emph{Wikipedia} and other sources on the Internet.
    \item Current events extracted from \emph{Wikipedia}~\cite{wiki_current}.
    \item Alternatively, we can manually construct a list of prominent events and extract
	  relevant information such as named entities, geographical location, and
	  time from ontologies such as: YAGO~\cite{yago}, Freebase~\cite{freebase}, etc.
  \end{itemize}  
  \item Diversifying and summarizing search events
  \begin{itemize} 
    \item Biographies of eminent personalities, for example United States presidents~\cite{wiki_potus}. 
    \item Historical timelines of various countries, for example for India~\cite{wiki_timeline}.
  \end{itemize}
\end{itemize}
		
\textbf{Structure}. Each event in our test bed is then composed of a fact with an accompanying 
query. Formally, a \emph{fact} in our testbed is a 4-tuple extracted from one of the
aforementioned sources:
$$
  \langle q,\mathcal{E}, g, t, \mathcal{W} \rangle
$$

where $q$  consists of keyword query describing the event, 
$\mathcal{E}$ is a bag of participating entities, 
$g$ is the geographic location, $t$ is the time of its 
occurrence, and $\mathcal{W}$ are important terms describing the 
event. 

\textbf{Metrics}. Based on the structure of the testbed of events, 
metrics such as \emph{precision}, \emph{recall} and \emph{F$_1$} 
can be utilized to measure the effectiveness of the algorithms for
detecting important events in semantically annotated corpora. 
How effectively the algorithm diversifies documents along multiple 
dimensions can be evaluated by metrics such as $\alpha$-nDCG~\cite{diverse}.
Quality of summaries can be measured by an 
automatic evaluation metric called \emph{Rouge}~\cite{rouge}. 
\section{Discussion}~\label{sec:discussion}

I briefly present some open technical challenges that I will address along
with the research objectives in my PhD dissertation.

\textbf{Mathematical Models}. One key aspect that occurs in the design
of the algorithms is that of computational models for named entities,
geographical locations and temporal expressions. What would be the most 
descriptive mathematical models for each of these semantic annotations?

\textbf{Similarity Functions}. Given a pair of named entities,
geographical locations or temporal expressions; 
how can we efficiently compute the similarity between the same type of
annotations?

\textbf{Efficiency \& Scalability}. Identifying data structures for indexing
corpora along with their semantic annotations, such that their asymptotic 
run times scale linearly with the size of the corpora.

\textbf{Evaluation}. Since evaluation of the solutions outlined are very subjective
in nature; what are other reliable sources of objective ground truth ? What other 
metrics can be employed to test the effectiveness of our methods ?
\section{Conclusion}~\label{sec:conclusion}

In this article I laid out an outline of the research 
work that I envisage to carry out for my PhD dissertation.
The research would in its culmination provide us methods 
to computationally extract world history as sequence of 
temporally ordered events and portray future events to 
take place from semantically annotated corpora. The research 
would also provide ways to perform semantic search and large
scale event analytics on these annotated corpora. I further
described already available resources that can be utilized 
for carrying out the research; test cases that can be built 
from encyclopedic resources on the Internet; and the metrics 
that can be utilized for evaluation.
\end{document}